\documentclass[structabstract]{aa}  
\usepackage{graphicx}
\usepackage{txfonts}
\usepackage{natbib}
\usepackage{longtable}
\begin{document}

\title{HIP 114328: a new refractory-poor and Li-poor solar twin
\thanks{Based on observations obtained at the
European Southern Observatory (ESO) Very Large Telescope (VLT) at Paranal Observatory, Chile (observing program 083.D-0871).}
}
\titlerunning{HIP 114328: a solar twin with low Li}

\newcommand{\teff}{$T_{\rm eff}$ }
\newcommand{\tsin}{$T_{\rm eff}$}
\newcommand{\tef}{T_\mathrm{eff}}
\newcommand{\logg}{\log g}
\newcommand{\feh}{\mathrm{[Fe/H]}}

\author{
Jorge Mel\'endez\inst{1} \and
Lucas Schirbel\inst{1} \and
TalaWanda R. Monroe\inst{1} \and
David Yong\inst{2} \and
Iv\'an Ram{\'{\i}}rez\inst{3} \and
Martin Asplund\inst{2}
}


\institute{
Departamento de Astronomia do IAG/USP, Universidade de S\~ao Paulo, Rua do Mat\~ao 1226, Cidade Universit\'aria, 
05508-900 S\~ao Paulo, SP, Brazil. e-mail:  jorge.melendez@iag.usp.br     
\and
Research School of Astronomy and Astrophysics,
The Australian National University, Cotter Road, Weston, ACT 2611, Australia
\and
McDonald Observatory and Department of Astronomy, University of Texas at Austin, USA
}

\date{Received ...; accepted ...}

 
  \abstract
   {The standard solar model fails to predict the very low lithium abundance in the Sun,
   which is much lower than the proto-solar nebula (as measured in meteorites). 
   This Li problem has been debated for decades, and it has been ascribed either to
   planet formation or to secular stellar depletion due to additional mixing below
   the convection zone, either during the pre-main sequence and thus
   possibly linked to planet formation or additionally on secular time-scales during the main sequence. 
   In order to test the evolution of Li, it is important to find solar twins in a range
   of ages, i.e., stars with about a solar mass and metallicity but in different evolutionary
   stages.  Also, the study of stars similar to the Sun is relevant 
   in relation to the signature of terrestrial planet formation around the Sun,
   and for anchoring photometric and spectroscopic stellar parameter scales.
   }
   {We aim to identify and analyse solar twins using high quality spectra, in order to study 
   Li depletion in the Sun and the possible relation between chemical abundance anomalies and planet formation.
}
   {We acquired high-resolution (R~$\sim$~110,000), high S/N ($\sim$300) ESO/VLT UVES 
   spectra of several solar twin candidates and the Sun (as reflected from the asteroid Juno). 
   Among the solar twin candidates we identify HIP 114328 as a solar twin and perform a 
   differential line-by-line abundance analysis of this star relative to the Sun.
   }
   {HIP 114328 has stellar parameters \teff = 5785$\pm$10 K, log $g$ = 4.38$\pm$0.03, 
   [Fe/H] = -0.022$\pm$0.009, and a microturbulent velocity 0.05$\pm$0.03 km s$^{-1}$ 
   higher than solar. The differential analysis shows that this star is chemically 
   very similar to the Sun. The refractory elements seem even slightly more depleted 
   than in the Sun, meaning that HIP 114328 may be as likely to form terrestrial planets
   as the Sun. HIP 114328 is about 2 Gyr older than the Sun, and is thus the second
   oldest solar twin analyzed at high precision. 
   It has a Li abundance of A(Li)$_{\rm NLTE}$ $\lesssim$ 0.46, which is about 4
times lower than in the Sun (A(Li)$_{\rm NLTE}$ = 1.07 dex), but close
to the oldest solar twin known, HIP 102152.
}
   {Based on the lower abundances of refractory elements when compared to other solar twins,
   HIP~114328 seems an excellent candidate to host rocky planets. 
   The low Li abundance of this star is consistent with its old age and fits very well 
   the emerging Li-age relation among solar twins of different ages.
}


\maketitle

%

\section{Introduction}

Since we can observe the Sun only at its current age, we have to rely upon younger and older 
stars to understand how the Sun would have been or how it will be at different evolutionary stages. 
The ideal sample of stars to compare the Sun with are solar twins \citep{cay96}.
Although solar twins are usually defined based on the similarity of their spectra to the Sun
 \citep[e.g.,][]{dat14} or stellar parameters \citep[e.g.,][]{ram09}, 
here we refer to solar twins as main sequence stars with about one-solar-mass and about 
solar composition, but spanning a range of ages. Having a mass and composition similar to the Sun ensures 
that solar twins will follow about the same evolutionary path as the Sun,
thus allowing us to study the evolution of the Sun in time.
As the stellar parameter space covered by one-solar-mass main sequence stars of roughly solar metallicity 
([Fe/H] = 0.0$\pm$0.1 dex) is broader than the working definition of solar twins given by \cite{ram09}, 
which is \teff within 100K, log $g$ within 0.1 dex and [Fe/H] within 0.1 dex of the solar values, 
all previous solar twins are included in our definition.

Of particular importance is the study of lithium; this element has been potentially related 
either to planet formation \citep{isr04,isr09,che06,gon10,tak10,del14,gon14} or to stellar depletion as stars 
evolve \citep{mel10,bau10,mon13}. A sample of solar twins with a range of ages is 
crucial to better understand this element, which can be used as an important constraint 
for non-standard stellar evolution models \citep{ct05, xd09, don09, bar10, den10, li12}.

Several bright (V $<$ 10) solar twins have been identified 
already \citep{por97,mel06,mel07,tak07,tak09,mel09,ram09,dat12,dat14,por14}, 
so they can be subject to high S/N, high resolving power ($R = \lambda/\delta \lambda$) studies, 
i.e. to high precision analyses using a high figure of merit $F =  (R [S/N])/\lambda$ \citep{nor01}. 
For example, the work by \cite{ram11} achieved a precision of about 0.01-0.02 dex in chemical 
abundances using $F \sim 4000$, while both \cite{mel12} and \cite{mon13} achieved a precision 
of about 0.005-0.010 dex with $F \sim 10 000$, and \cite{mel14} achieved a precision 
of about 0.005 dex with $F \sim 15 000$.
Among those solar twins studied at high precision (F $\gtrsim$4000), 
only 16 Cyg B \citep{ram11} and HIP  102152 \citep{mon13} seem older than the Sun.

In this Letter we report the identification of another solar twin older than the Sun, 
HIP 114328 (HD 218544), thus bringing important insights on the evolution of Li and therefore 
on the mechanisms that destroy this fragile element in solar type stars. Also, we will discuss the refractory-poor
abundance pattern of this star in the context of chemical anomalies and planet formation.

\section{Observations}

Based on their colors and Hipparcos parallaxes, we selected eight solar twin candidates 
for spectroscopic observations, HIP1536, HIP 3238, HIP 10725, HIP11514, HIP 106288, 
HIP 109381, HIP 114328 and HIP 117499, 
as well as the asteroid Juno to obtain a reference solar spectrum. 
The observations were taken using
UVES in dichroic mode, with the 346nm setting (306-387 nm) in the blue arm and the 
580-nm setting (480 - 682 nm) in the red arm. 

Most spectral lines used are in the red arm, where we achieved 
R = 110 000 using the 0.3 arcsec slit. The typical S/N is about 285 per pixel,
thus our figure of merit is F $\sim$ 5000.
In the UV we used a slit of 0.6 arcsec, resulting in R = 65 000.

\begin{figure}
\resizebox{\hsize}{!}{\includegraphics{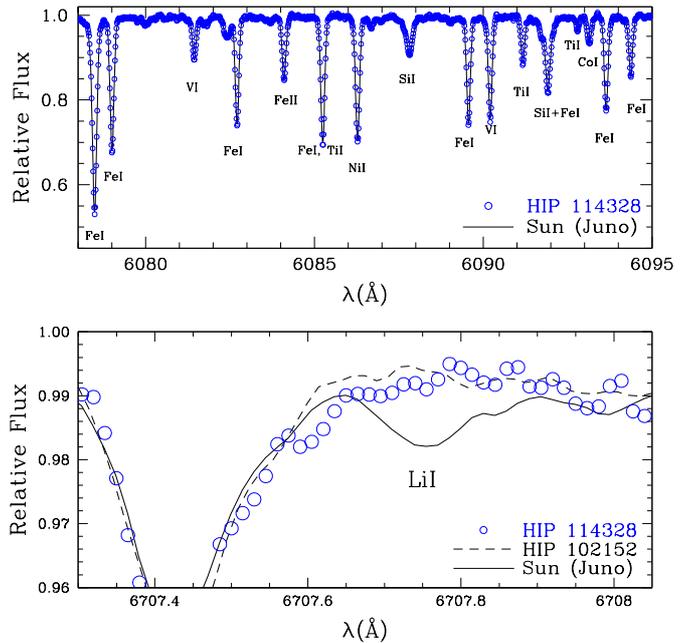}}
\caption{
Comparison of the spectra of HIP 114328 (blue circles) and the Sun (solid line)
around 609 nm (top panel), showing that both stars have similar spectra.
The Li feature for HIP 114328, HIP 102152 and the Sun, is shown in the bottom panel.
HIP 114328 has a Li feature substantially weaker than solar and comparable
to the old solar twin HIP 102152.
}
\label{spectra}
\end{figure}

The spectral orders were extracted and wavelength calibrated using IRAF. Further data processing 
was performed with IDL. A comparison of the solar twin candidates to the Sun, revealed that 
only the spectrum of HIP 114328 matched well the solar spectrum, hence this star was selected 
for a further detailed analysis.
Part of the reduced spectra of HIP 114328 and the Sun is shown
in Fig.~\ref{spectra} in the region 6078-6095 \AA\ and around
the Li feature. The spectra are very similar, except for the
Li feature, with HIP 114328 showing a much weaker feature than the Sun,
and similar to the old solar twin HIP 102152 \citep{mon13}.

\section{Abundance analysis}

The analysis is similar to that presented in \cite{mel12,mel14} and \cite{mon13}. 
The main difference is that now all equivalent width (EW) measurements were performed by hand, 
instead of first having a set of automatic measurements
with ARES \citep{sou07}. 
The number of outliers in the hand measurements is significantly smaller than 
those obtained in our previous works when using EWs measured automatically. 
Thus, only for a few lines we needed to check the manual measurements.
The line list is from \cite{mel14}, and is an 
extended version of that presented in \cite{mel12}.

The same differential approach as in our previous papers was used to obtain 
stellar parameters and chemical abundances, i.e., we followed a strictly differential
line-by-line analysis. 
We adopted ATLAS9 model atmospheres \citep{cas04}, although the differential analysis 
of solar twins is essentially insensitive to the chosen grid of model atmospheres \citep{mel12}. 
The analysis was performed using the 2002 version of MOOG
\citep{sne73}. As shown in \cite{mel12,mel14} and \cite{mon13}, differential NLTE corrections 
in solar twins are negligible, hence they are not taken into account here.

The differential spectroscopic equilibrium of HIP 114328 relative to the Sun 
results in stellar parameters of \teff = 5785$\pm$10 K ($\Delta$\teff = +8$\pm$10 K), log $g$ = 4.38$\pm$0.03 dex
($\Delta$log $g$ = $-$0.06$\pm$0.03 dex), 
[Fe/H] = $-$0.022$\pm$0.009 dex, and a microturbulent velocity +0.05$\pm$0.03 km s$^{-1}$ higher than solar.
The errors in the stellar parameters were estimated based on the observational
uncertainties and take into account the degeneracy in the stellar parameters.

Once the stellar parameters were set, we computed differential abundances
using the measured EWs, except for Li, which was analysed by spectrum synthesis
using the line list of \cite{mel12}. Hyperfine structure was taken into account
for V, Mn, Co and Cu.
The differential abundances are provided in Table \ref{abund},
as well as the observational errors (standard errors), the errors due to uncertainties in the stellar parameters,
and the total error, obtained by adding in quadrature the observational and systematic errors.

\begin{figure}
\resizebox{\hsize}{!}{\includegraphics{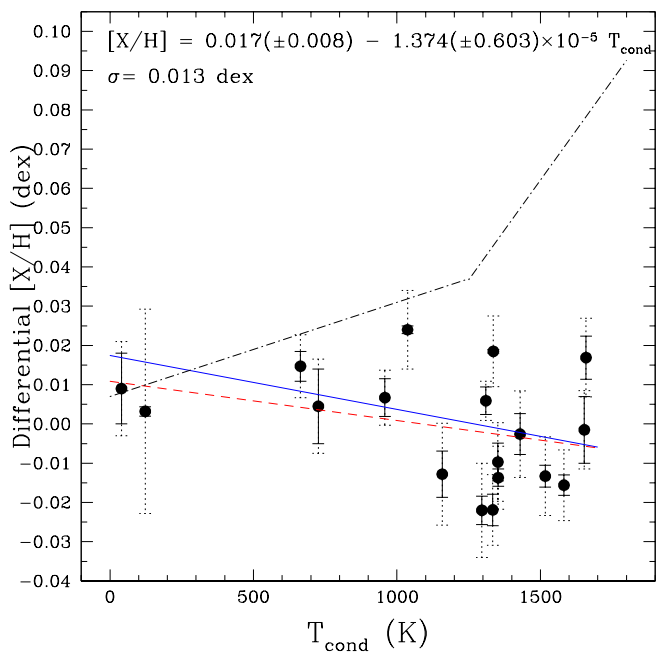}}
\caption{
Abundance pattern of HIP 114328 (circles) versus condensation temperature.
The solid error bars are based only on the observational uncertainties, 
and the dotted error bars are the total errors (average $\sim$0.011 dex).
The solid line is a fit taking into account the total error bars
and the dashed line is a fit without considering error bars. 
The element-to-element scatter from the fit is only $\sigma$ = 0.013 dex,
in good agreement with the mean error bar.
The dot-dashed line is  the mean abundance pattern of 11 solar twins in
\cite{mel09}, shifted vertically to match the volatile element carbon.
}
\label{tcond}
\end{figure}

In Fig.~\ref{tcond} we plot the differential abundances [X/H] between HIP 114328 and the Sun
(circles) as a function of equilibrium condensation temperature \citep[T$_{\rm cond}$,][]{lod03}. 
The abundance pattern of HIP 114328 is similar to solar and seems
slightly more depleted in refractories than the Sun, as shown
by the fits (solid and dashed lines). Considering the uncertainties, 
the refractory-to-volatile ratio in HIP 114328 is similar to solar.
For comparison, the mean abundance pattern of eleven solar twins 
\citep{mel09}, is shown by a dot-dashed line,
showing that HIP 114328 is indeed depleted in refractories.

The element-to-element scatter (0.013 dex) of the differential
abundances in HIP 114328 is similar to the typical 
error bar of the differential abundances (0.011 dex), showing that our
error bars are realistic and that we achieved a typical
error of $\sim$0.01 dex.
 
The LTE lithium abundance in HIP 114328 is A(Li) $\lesssim$ 0.41$\pm$0.13,
and the NLTE abundance is A(Li)$_{\rm NLTE}$ $\lesssim$ 0.46$\pm$0.13, 
using the NLTE corrections of \cite{lin09}. This low Li abundance is
slightly higher than the Li abundance in the old solar twin HIP 102152
\citep{mon13}, and as discussed below, reinforces the idea of
secular stellar depletion.

\section{Discussion}

HIP 114328 has been little studied in the literature, with only two papers
reported by SIMBAD, one on stellar activity \citep{jen11} and the other
one on a list of candidates for targeted transit searches \citep{her12}.
There are three earlier papers on delta Scuti stars but they are actually 
misidentifications, and refer to DY Peg (HD 218549) and not to
HIP 114328 (HD 218544).

The abundance pattern of HIP 114328 is very similar to solar (Fig. \ref{tcond})
and seems even slightly more depleted in refractories than the Sun.
The abundance pattern of the Sun is different from most solar
twins \citep{mel09,ram09}, probably due to the formation of the
terrestrial planets in the solar system \citep{cha10}.
Thus, the chemical similarity between HIP 114328 and the Sun,
may suggest that HIP 114328 was as well equipped as the Sun to host rocky planets.

An alternative hypothesis to explain the Sun's abundances anomalies is that the viewing angle 
of solar twins from Earth is different than the angle of the Sun when observed from Earth. However, 
a detailed analysis of solar spectra taken at different solar latitudes revealed no abundance differences \citep{kis11}.
Another explanation put forward by \cite{one11}, is that the lack of refractory elements may reflect
actually that the star was born in a dense environment, as suggested
by the analysis of one solar twin in the open cluster M67.
Interestingly, the recent work by \cite{adi14}, shows that older stars have a lower refractory-to-volatile 
ratio than younger stars, thus suggesting that age may play a role in the trends with condensation temperature.
However, the recent discovery of clear abundance differences between the
binary components of 16 Cygni \citep{law01,ram11,tuc14}, where the secondary hosts a giant planet
but no planet has been detected around the primary despite more than two decades of
radial velocity monitoring,
strongly suggests that planet formation can indeed imprint chemical signatures
on the composition of their host stars. 
Notice that although both \cite{sch11} and \cite{tak05a} found no abundance 
difference between 16 Cyg A and B, their work is based on spectra with a lower 
figure of merit (i.e., lower quality) than in \cite{tuc14}, who used a resolving power 
R = 81 000 and S/N = 700, implying $F$ = 9450 at 6000 \AA. On the other hand, 
\cite{sch11} used R = 45 000 and S/N=750, hence their $F$ = 5625. 
The work of \cite{tak05a}, made use of R = 70 000 spectra, which had a low S/N = 90 - 130 \citep{tak05b}, 
resulting in a much lower figure of merit ($F$ = 1280).

Using our precise stellar parameters with their error bars, Yonsei-Yale isochrones
\citep{kim02,dem04} and probability distribution functions \citep[as described in][]{mel12},
we estimate an age and mass for HIP 114328 of 6.7$^{+0.6}_{-1.1}$ Gyr
and 0.99$\pm$0.01 M$_\odot$, respectively. The old age of this solar twin
is consistent with the low activity level measured by \cite{jen11},
$R_{HK}$ = -5.024, fitting well the activity-age relation of solar twins \citep{ram14}.

\begin{figure}
\includegraphics[width=9.5cm]{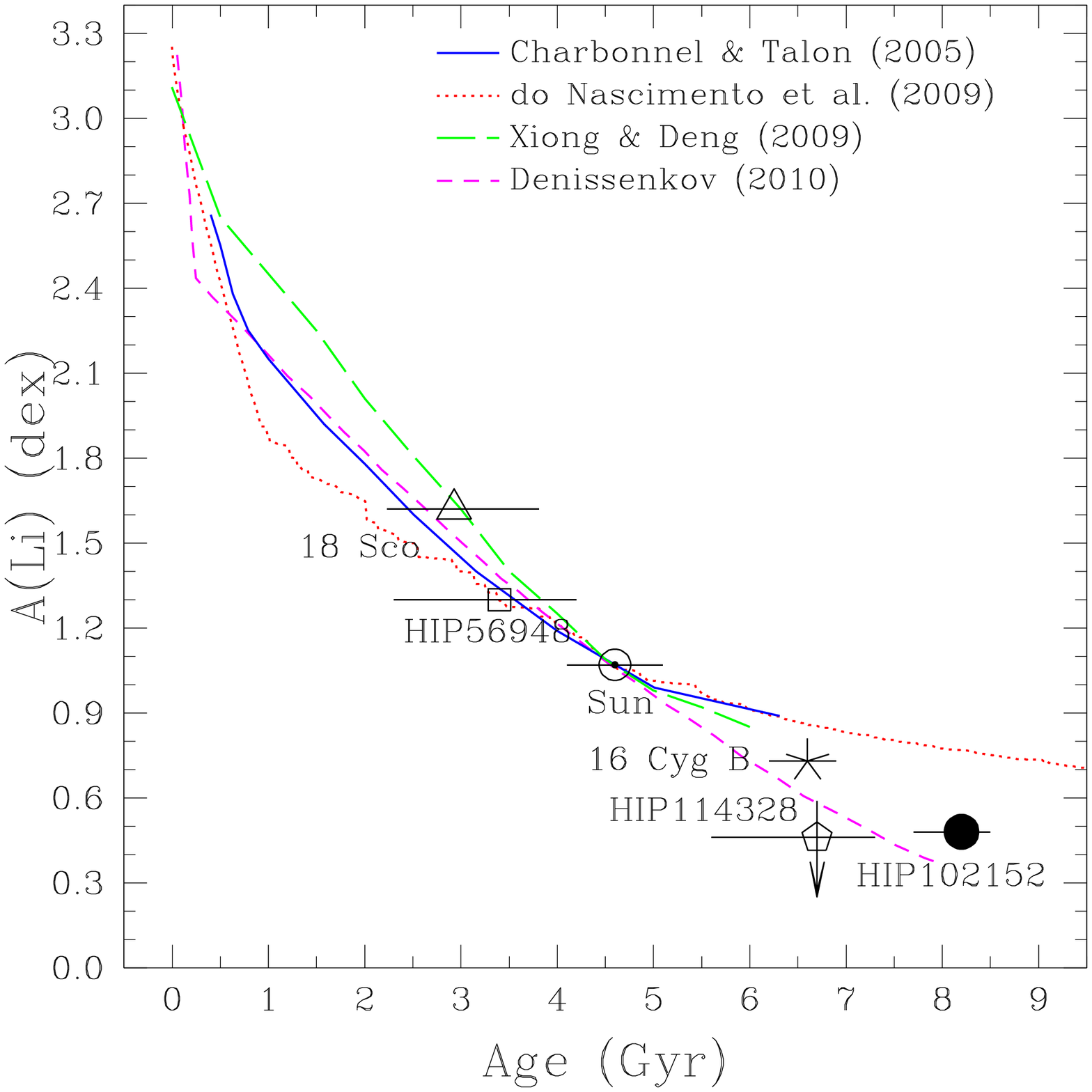}
\caption{NLTE Li abundances vs. age for the Sun and solar twins observed at
high spectral resolution and high S/N.
The total error bar ($\pm \sigma$) of the Li abundance is about the size of 
the symbols, while the error bars in age are shown by horizontal lines.
For comparison we show the models by \cite{ct05,don09,xd09,den10}, shifted 
to reproduce our observed NLTE solar Li abundance. The model with 
initial rotation velocity of 50 km s$^{-1}$ was adopted for \cite{ct05}.
HIP 114328 is shown by a pentagon, and helps to define a clear Li-age correlation.
}
\label{age}
\end{figure}

We plot the Li abundance and age of HIP 114328 in Fig. \ref{age},
together with the solar twins used by \cite{mon13}, which are all
based on analyses with high figure of merit (F $\geq$ 4000) and
R $\geq$ 60 000. In this plot we updated the age of 16 Cyg B \citep[previously from][]{ram11}
using the most precise stellar parameters of Tucci-Maia et al. (2014), 
resulting in an age of 6.6$^{+0.3}_{-0.4}$ Gyr.
We also plot several theoretical tracks of non-standard models
of lithium depletion \citep{ct05,don09,xd09,den10}.
HIP 114328 fits well the Li-age correlation
found by \cite{mon13}. This connection between Li and age has been already suggested, 
albeit with larger uncertainties, in our earlier works \citep{mel10,bau10}.
This reinforces that stellar Li depletion is secular and not related
to planet formation \cite[e.g.,][]{isr09}; 
the low Li content in the Sun is perfectly normal for its age.
Overall there is a good agreement with all non-standard
models shown in Fig. \ref{age}.

\section{Conclusions}

We achieve a precision of about 0.01 dex in the differential
analysis of HIP 114328 relative to the Sun.
This solar twin has a chemical composition similar to solar, hence
it is a good candidate to look for potential rocky planets. 
First identifying a Sun 2.0, or solar twin, with potential terrestrial planet formation, 
such as HIP 114328, is perhaps a good strategy for workers to consider as they search 
for an Earth 2.0, or Earth-sized planet in the habitable zone of a sun-like star.  
Although originally not included in our HARPS planet survey
around solar twins \citep[][ESO Large Programme 188.C-0265, PI: J. Mel\'endez]{ram14},
we will start to search for planets around this star.

We determine an old age for HIP 114328 ($\sim$7 Gyr), which makes it
important to study the depletion of Li with age. The low Li abundance
of HIP 114328 fits very well an emerging tight relation between 
Li and age, and shows that Li could be used as a cosmochronometer,
thus helping to derive ages in main sequence stars \citep{don09,li12}. 
The study of more solar twins in a range of ages will help to
better constrain non-standard stellar evolution models.

\begin{acknowledgements}

J.M. would like to acknowledge support from
FAPESP (2012/24392-2) and CNPq ({\em Bolsa de Produtividade}).
T.M. and L.S. acknowledge support from FAPESP (2010/19810-4 and 2013/25008-4).
M.A and D.Y acknowledge support from the Australian Research Council (grants FL110100012 and DP120100991).
\end{acknowledgements}

\clearpage

\begin{table}
\caption{Differential abundances of HIP 114328 relative to the Sun and errors.}
\label{abund}
\centering 
\renewcommand{\footnoterule}{}  
\begin{tabular}{lrrrrrrrrrr} 
\hline    
\hline 
{Element}& LTE   & $\Delta \tef$ & $\Delta$log $g$ & $\Delta v_t$ & $\Delta$[Fe/H] & param\tablefootmark{a} & obs\tablefootmark{b} & total\tablefootmark{c} \\
{}       &       &   +10K           &  +0.03 dex      & +0.03 km s$^{-1}$  & +0.01 dex   &  &  &  \\
{}       & (dex) & (dex) & (dex)         & (dex)           & (dex)       & (dex)        & (dex) & (dex) \\
\hline
C  &  0.009 & -0.005 &  0.007 &  0.000 & -0.001 & 0.009 & 0.009 & 0.012\\  
N  &  0.003 &  0.013 &  0.007 &  0.020 &  0.009 & 0.026 & 0.001 & 0.026\\ 
Na &  0.007 &  0.005 & -0.001 &  0.000 &  0.001 & 0.005 & 0.005 & 0.007\\  
Mg &  0.019 &  0.005 & -0.006 & -0.004 &  0.000 & 0.009 & 0.001 & 0.009\\  
Al & -0.002 &  0.005 & -0.001 & -0.001 &  0.001 & 0.005 & 0.009 & 0.010\\  
Si &  0.006 &  0.002 &  0.001 & -0.001 &  0.001 & 0.003 & 0.004 & 0.005\\  
S  &  0.015 & -0.004 &  0.006 &  0.000 &  0.000 & 0.007 & 0.004 & 0.008\\    
Ca & -0.013 &  0.007 & -0.005 & -0.005 &  0.001 & 0.010 & 0.003 & 0.010\\  
Sc &  0.017 &  0.008 &  0.002 & -0.001 &  0.001 & 0.008 & 0.006 & 0.010\\  
Ti & -0.016 &  0.004 &  0.005 & -0.005 & -0.002 & 0.008 & 0.003 & 0.009\\  
V  & -0.003 &  0.010 &  0.001 & -0.002 &  0.001 & 0.010 & 0.005 & 0.011\\  
Cr & -0.022 & -0.002 &  0.010 & -0.004 &  0.002 & 0.011 & 0.004 & 0.012\\  
Mn & -0.013 &  0.008 & -0.004 & -0.007 &  0.002 & 0.012 & 0.006 & 0.013\\  
Fe & -0.022 &  0.001 &  0.005 & -0.006 &  0.002 & 0.008 & 0.004 & 0.009\\  
Co & -0.010 &  0.008 &  0.003 & -0.002 &  0.001 & 0.009 & 0.005 & 0.010\\  
Ni & -0.014 &  0.006 &  0.000 & -0.005 &  0.002 & 0.008 & 0.002 & 0.008\\  
Cu &  0.024 &  0.006 & -0.001 & -0.007 &  0.003 & 0.010 & 0.001 & 0.010\\  
Zn &  0.005 &  0.001 &  0.002 & -0.006 &  0.003 & 0.007 & 0.010 & 0.012\\    
\hline       
\end{tabular}
\tablefoot{Abundances of V, Mn, Co, and Cu account for HFS.\\
\tablefoottext{a}{Adding errors in stellar parameters}
\tablefoottext{b}{Observational errors}
\tablefoottext{c}{Total error (stellar parameters and observational)}
}
\end{table}


\begin{thebibliography}{}
\bibitem[Adibekyan et al.(2014)]{adi14} Adibekyan, V.~Z., Gonz{\'a}lez Hern{\'a}ndez, J.~I., Delgado Mena, E., et al.\ 2014, \aap, 564, L15 

\bibitem[Baraffe \& Chabrier(2010)]{bar10} Baraffe, I., \& Chabrier, G.\ 2010, \aap, 521, A44 

\bibitem[Baumann et al.(2010)]{bau10} Baumann, P., Ram{\'{\i}}rez, I., Mel{\'e}ndez, J., Asplund, M., \& Lind, K.\ 2010, \aap, 519, A87 

\bibitem[Castelli 
\& Kurucz(2004)]{cas04} Castelli, F., \& Kurucz, R.~L.\ 2004, arXiv:astro-ph/0405087 

\bibitem[Cayrel de Strobel(1996)]{cay96} Cayrel de Strobel, G.\ 1996, \aapr, 7, 243

\bibitem[Chambers(2010)]{cha10} Chambers, J.~E.\ 2010, \apj, 724, 92 

\bibitem[Charbonnel \& Talon(2005)]{ct05} Charbonnel, C., \& Talon, S.\ 2005, Science, 309, 2189 

\bibitem[Chen 
\& Zhao(2006)]{che06} Chen, Y.~Q., \& Zhao, G.\ 2006, \aj, 131, 1816 

\bibitem[Datson et al.(2012)]{dat12} Datson, J., Flynn, C., \& Portinari, L.\ 2012, \mnras, 426, 484 

\bibitem[Datson et al.(2014)]{dat14} Datson, J., Flynn, C., 
\& Portinari, L.\ 2014, \mnras, 194 

\bibitem[Delgado Mena et 
al.(2014)]{del14} Delgado Mena, E., Israelian, G., Gonz{\'a}lez Hern{\'a}ndez, J.~I., et al.\ 2014, \aap, 562, A92 

\bibitem[Demarque et al.(2004)]{dem04} Demarque, P., Woo, J.-H., Kim, Y.-C., \& Yi, S.~K.\ 2004, \apjs, 155, 667

\bibitem[Denissenkov(2010)]{den10} Denissenkov, P.~A.\ 2010, \apj, 719, 28 

\bibitem[do Nascimento et al.(2009)]{don09} Do Nascimento, J.~D., Jr., Castro, M., Mel{\'e}ndez, J., Bazot, M., Th{\'e}ado, S., Porto de Mello, G.~F., \& de Medeiros, J.~R.\ 2009, \aap, 501, 687 

\bibitem[Gonzalez et al.(2010)]{gon10} Gonzalez, G., Carlson, M.~K., \& Tobin, R.~W.\ 2010, \mnras, 407, 314 

\bibitem[Gonzalez(2014)]{gon14} Gonzalez, G.\ 2014, MNRAS, in press, arXiv:1404.0574 

\bibitem[Herrero et al.(2012)]{her12} Herrero, E., Ribas, I., Jordi, C., Guinan, E.~F., \& Engle, S.~G.\ 2012, \aap, 537, A147 

\bibitem[Israelian et al.(2004)]{isr04} Israelian, G., Santos, N.~C., Mayor, M., \& Rebolo, R.\ 2004, \aap, 414, 601 

\bibitem[Israelian et al.(2009)]{isr09} Israelian, G., 
Delgado Mena, E., Santos, N.~C., et al.\ 2009, \nat, 462, 189 

\bibitem[Jenkins et al.(2011)]{jen11} Jenkins, J.~S., Murgas, F., Rojo, P., et al.\ 2011, \aap, 531, A8 

\bibitem[Kim et al.(2002)]{kim02} Kim, Y.-C., Demarque, P., Yi, S.~K., \& Alexander, D.~R.\ 2002, \apjs, 143, 499 

\bibitem[Kiselman et 
al.(2011)]{kis11} Kiselman, D., Pereira, T.~M.~D., Gustafsson, B., et al.\ 2011, \aap, 535, A14 

\bibitem[Laws \& Gonzalez(2001)]{law01} Laws C., \& Gonzalez G.\ 2001, \apj, 553, 405 

\bibitem[Li et al.(2012)]{li12} Li, T.~D., Bi, S.~L., Chen, Y.~Q., et al.\ 2012, \apj, 746, 143 

\bibitem[Lind et 
al.(2009)]{lin09} Lind, K., Asplund, M., \& Barklem, P.~S.\ 2009, \aap, 503, 541 

\bibitem[Lodders(2003)]{lod03} Lodders, K.\ 2003, \apj, 591, 1220 

\bibitem[Mel\'{e}ndez et al.(2006)]{mel06} Mel\'{e}ndez, J., Dodds-Eden, K. \& Robles, J. A.  2006, \apj, 641, L133

\bibitem[Mel\'{e}ndez \& Ram\'{i}rez(2007)]{mel07} Mel\'{e}ndez, J. \& Ram\'{i}rez, I.  2007, \apj, 669, L89

\bibitem[Mel{\'e}ndez et al.(2009)]{mel09} Mel{\'e}ndez, J., Asplund, M., Gustafsson, B., \& Yong, D.\ 2009, \apjl, 704, L66

\bibitem[Mel{\'e}ndez et al.(2010)]{mel10} Mel{\'e}ndez, J., et al.\ 2010, \apss, 328, 193

\bibitem[Mel{\'e}ndez et 
al.(2012)]{mel12} Mel{\'e}ndez, J., Bergemann, M., Cohen, J.~G., et al.\ 2012, \aap, 543, A29 

\bibitem[Mel{\'e}ndez et al.(2014)]{mel14} Mel{\'e}ndez, J.,  Ram\'{i}rez, I., Karakas, A. et al.\ 2014, \apj, submitted 

\bibitem[Monroe et al.(2013)]{mon13} Monroe, T.~R., 
Mel{\'e}ndez, J., Ram{\'{\i}}rez, I., et al.\ 2013, \apjl, 774, L32 

\bibitem[Norris et al.(2001)]{nor01} Norris, J.~E., Ryan, 
S.~G., \& Beers, T.~C.\ 2001, \apj, 561, 1034 

\bibitem[{\"O}nehag et al.(2011)]{one11} {\"O}nehag, A., Korn, A., Gustafsson, B., Stempels, E., \& Vandenberg, D.~A.\ 2011, \aap, 528, A85

\bibitem[Porto de Mello \& da Silva(1997)]{por97} Porto de Mello, G.~F., \& da Silva, L.\ 1997, \apjl, 482, L89

\bibitem[Porto de Mello et 
al.(2014)]{por14} Porto de Mello, G.~F., da Silva, R., da Silva, L., \& de Nader, R.~V.\ 2014, \aap, 563, A52 

\bibitem[Ram{\'{\i}}rez et al.(2009)]{ram09} Ram{\'{\i}}rez, I., Mel{\'e}ndez, J., \& Asplund, M.\ 2009, \aap, 508, L17 

\bibitem[Ram{\'{\i}}rez et al.(2011)]{ram11} Ram{\'{\i}}rez, 
I., Mel{\'e}ndez, J., Cornejo, D., Roederer, I.~U., \& Fish, J.~R.\ 2011, \apj, 740, 76 

\bibitem[Ram{\'{\i}}rez et al.(2014)]{ram14} Ram{\'{\i}}rez, 
I., Mel{\'e}ndez, J., Bean, J. et al.\ 2014, submitted

\bibitem[Schuler et al.(2011)]{sch11} Schuler, S.~C., Cunha, 
K., Smith, V.~V., et al.\ 2011, \apjl, 737, L32 

\bibitem[Sneden(1973)]{sne73} Sneden, C.~A.\ 1973, Ph.D.~Thesis,  

\bibitem[Sousa et al.(2007)]{sou07} Sousa, S.~G., Santos, N.~C., Israelian, G., Mayor, M., \& Monteiro, M.~J.~P.~F.~G.\ 2007, \aap, 469, 783 

\bibitem[Takeda(2005)]{tak05a} Takeda, Y.\ 2005, \pasj, 57, 83

\bibitem[Takeda et al.(2005)]{tak05b} Takeda, Y., Sato, B., 
Kambe, E., et al.\ 2005, \pasj, 57, 13

\bibitem[Takeda et al.(2007)]{tak07} Takeda, Y., Kawanomoto, S., Honda, S., Ando, H., \& Sakurai, T.\ 2007, \aap, 468, 663 

\bibitem[Takeda \& Tajitsu(2009)]{tak09} Takeda, Y., \& Tajitsu, A.\ 2009, \pasj, 61, 471 

\bibitem[Takeda et al.(2010)]{tak10} Takeda, Y., Honda, S., Kawanomoto, S., Ando, H., \& Sakurai, T.\ 2010, \aap, 515, A93 

\bibitem[Tucci Maia et al.(2014)]{tuc14} Tucci Maia, M, Mel{\'e}ndez, J., \&  Ram\'{i}rez, I.\ 2014, \apj ~Letters, submitted 

\bibitem[Xiong \& Deng(2009)]{xd09} Xiong, D.~R., \& Deng, L.\ 2009, \mnras, 395, 2013 


\end{thebibliography}
\end{document}